# Spin dynamics, electronic and thermal transport properties of two-dimensional CrPS$_4$ single crystal

Q. L. Pei[1], X. Luo[1*], G.T. Lin[1], J. Y. Song[1], L. Hu[1], Y. M. Zou[2], L. Yu[2], W. Tong[2], W. H. Song[1], W. J. Lu[1] and Y. P. Sun[2,1,3*]

[1] Key Laboratory of Materials Physics, Institute of Solid State Physics, Chinese Academy of Sciences, Hefei, 230031, China

[2] High Magnetic Field Laboratory, Chinese Academy of Sciences, Hefei, 230031, China

[3] Collaborative Innovation Center of Advanced Microstructures, Nanjing University, Nanjing, 210093, China

## Abstract

2-Dimensional (2D) CrPS$_4$ single crystals have been grown by the chemical vapor transport method. The crystallographic, magnetic, electronic and thermal transport properties of the single crystals were investigated by the room-temperature X-ray diffraction, electrical resistivity $\rho(T)$, specific heat $C_P(T)$ and the electronic spin response (ESR) measurements. CrPS$_4$ crystals crystallize into a monoclinic structure. The electrical resistivity $\rho(T)$ shows a semiconducting behavior with an energy gap $E_a$=0.166 eV. The antiferromagntic (AFM) transition temperature is about $T_N$=36 K. The spin flipping induced by the applied magnetic field is observed along the $c$ axis. The magnetic phase diagram of CrPS$_4$ single crystal has been discussed. The extracted magnetic entropy at $T_N$ is about 10.8 J/mol K, which is consistent with the theoretical value $R\ ln(2S + 1)$ for $S = 3/2$ of the Cr$^{3+}$ ion. Based on the mean-field theory, the magnetic exchange constants $J_1$ and $J_c$ corresponding to the interactions of the intralayer and between layers are about 0.143 *meV* and -0.955 *meV* are obtained based on the fitting of the susceptibility above $T_N$, which agree with the results obtained from the ESR measurements. With the help of the strain for tuning the magnetic properties, monolayer CrPS$_4$ may be a promising candidate to explore 2D magnetic semiconductors.

Corresponding author: xluo@issp.ac.cn and ypsun@issp.ac.cn

# I Introduction

Two-dimensional (2D) materials have been attracted much attention due to their highly tunable physical properties and immense potential in scalable device applications.[1–4] Especially, the emergence of transition-metal dichalcogenides (TMD) is particularly advantageous as it opens the door to many physical properties not available in grapheme.[5-7] For example, in WTe$_2$ the large magnetoresistance without saturation at the high magnetic field leads to a new direction in the study of magnetoresistivity which might be useful for spintronic applications.[7-8] Another interesting possibility brought by transition-metal elements is the magnetism, which has been investigated to a lesser extent in the current 2D compounds.[9] In this regard, transition-metal tri- or tetra- chalcogenides, such as MnPS$_3$ and CrPS$_4$, represent a rather attractive material family. Similar to dichalcogenides, these compounds also show 2D characteristic with weak interlayer van der Waals interactions. Furthermore, these materials are known to exhibit a large variety of magnetic phases, making them ideal candidates for exfoliated 2D magnets.[10–12]

The layered CrPS$_4$ crystallizes into a monoclinic symmetry.[13,14] The sulfur layers are hexagonally close-packed, the Cr atoms are localized in the octahedral holes of S atoms and P atoms are surrounded by four S atoms in tetrahedral coordination. The van der Waals gaps exist between the sulfur layers along the *c* axis and the cleaving surface is parallel to the *ab* plane. The Cr-based compound CrPS$_4$ was synthesized thirty years ago, and an antiferromagnetic (AFM) temperature $T_N$=36 K of the insulating CrPS$_4$ was just reported.[14] To explore the possible exfoliated 2D magnet in CrPS$_4$, the motivation of the present paper aims to gain a better understanding of the bulk properties of CrPS$_4$ single crystal so that future studies on single-layer CrPS$_4$ will have a firmer foundation. Herein, we did the research by the magnetization, electronic, thermal transport and the electronic spin response (ESR) measurements on the layered CrPS$_4$ single crystal.

# II Experimental results

Single crystalline specimens of CrPS$_4$ were prepared by the chemical vapor transport method. Cr powders (99.99 %, Alfa Aesar), red P powders (99.99 %, Alfa Aesar) and S powders (99.99 %, Alfa Aesar) with mole ratio 1:1:4 were weighted and loaded into a silicon quartz tube, which was sealed under high vacuum. All were done in an Ar-filled glove box. The sealed quartz tubes were put in a two-zone tube furnace. The hot side is about 650 ºC while the cold side is about 550 ºC, and dwelled for 7 days, then slowly cooled down to room temperature with 100ºC per hour. Hexagonal shape single crystals with shining surface were observed. The size of the crystal was about 4*6*0.5 mm$^3$. The single crystals were air-stable and soft and can be easily exfoliated. Powder X-ray diffraction (XRD) patterns were taken with Cu $K_{\alpha 1}$ radiation (λ=0.15406 nm) using a PANalytical X'pert diffractometer at room temperature. The magnetic properties were carried out by the magnetic property measurement system (MPMS-XL5). The electrical transport measurements were performed by a four-probe method to eliminate the contact resistance. The measurement of specific heat was carried out by a heat-pulse relaxation method on Physical Properties Measurement System (PPMS-9T). The ESR measurement of the CrPS$_4$ single crystal was performed using a Bruker EMX plus model spectrometer operating at X-band frequencies (9.4 GHz) at different temperatures.

## III Results and discussion

The CrPS$_4$ crystallizes into a layered structure. As shown in the inset of Fig. 1 (a), hexagonal-close-packed sulfur layers parallel to the *bc* plane are the fundamental basis of the crystal structure and the Cr and P atoms are in distorted octahedral and tetrahedral interstices, respectively. The powdered XRD data were collected on crushed single crystals of CrPS$_4$ at room temperature as shown in Fig. 1 (a). It also presents the structural Rietveld refinement profiles of the XRD data using FullProf software.[15] The refinements of the XRD data indicate that the crystals are single-phase since no extra peaks were observed. The fitted parameters are shown in the Table I. The flat surface was identified as the (001) planes, as shown in the Fig. 1 (b) and the crystal used is presented in the inset of Fig. 1 (b).

In order to investigate the macroscopic magnetic properties of the CrPS$_4$ single crystal, we carried out the measurement of the magnetization as the function of temperature and magnetic field. Figure 2 shows the temperature dependence of magnetization (*M(T)*) under the zero-field-cooled (ZFC) and field-cooled (FC) modes with the applied magnetic field H= 0.01 T. Fig. 2 (a) and (b) show the *M(T)* with *H* parallel and perpendicular to the *c* axis, respectively. The Néel temperature $T_N$ is observed around 36 K, which is consistent with the reported data.[14] The magnetization below the $T_N$ is anisotropic with $M_{ab} > M_c$, which indicates that the easy magnetization axis is along the direction of the $c$ axis.[16] In the paramagnetic (PM) state, the temperature dependence of susceptibility ($\chi(T)$) follows the Curie-Weiss law shown by solid lines in insets of Fig.2 (a) and (b). The Curie-Weiss law for the magnetic susceptibility $\chi$ is described as follows:

$$\chi(T) = \frac{M}{H} = \frac{C}{T - \theta_P} + \chi_0 \tag{1}$$

Where $C$ is the Curie constant, $\theta_P$ is the Weiss temperature and $\chi_0$ is the Pauli PM constant. From the least-square analysis by Eq. (1), we can obtain the Weiss temperature $\theta_P(H /\!/ c) = \theta_P(H \perp c) =$ 19.4 K and Curie constant $C(H /\!/ c) = 4.23$ emu K/mol, $C(H \perp c) = 4.13$ emu K/mol, respectively.

The effective magnetic moment $\mu_{eff}$ can be obtained by:

$$\mu_{eff}/\mu_B = \sqrt{3k_B C / N_A Z} \tag{2a}$$

where $k_B = 1.38 \times 10^{-16}$ erg/K, $N_A = 6.02 \times 10^{23}$ mol$^{-1}$ and $Z$ is the atom number per unit cell. The value of $\mu_{eff}$ can be simplified:

$$\mu_{eff} = \sqrt{8C} \mu_B \tag{2b}$$

From the Eq. (2b), we can get the effective magnetic moment $\mu_{eff}$

$$\mu_{eff} = \sqrt{8C} \mu_B = 5.8 \mu_B \quad (H /\!/ c) \tag{2c}$$

$$\mu_{eff} = \sqrt{8C} \mu_B = 5.74 \mu_B \quad (H \perp c) \tag{2d}$$

The theoretical value of $Cr^{3+}$ ion can be estimated by:

$$\mu_{eff} = g\sqrt{S(S+1)} \mu_B = 3.87 \mu_B \tag{3}$$

where the value of $g$ is 2. The experimental values of $\mu_{eff}$ are larger than the theoretical value of $Cr^{3+}$ ion, which may be related to the strong spin-lattice coupling existing above the $T_N$ in 2D $CrPS_4$ single crystal.[1]

In order to further investigate the nature of the AFM structure of $CrPS_4$ single crystal, the magnetic field dependence of the magnetization ($M(H)$) for $CrPS_4$ single crystal with different directions at $T=5$ K are shown in Fig. 3 (a) and (b), respectively. Interestingly, a magnetic switching effect is observed when the magnetic field is along the $c$ axis, as shown in Fig. 3 (a). The critical field $H_C$ is about 0.9 T at $T=5$ K. As the measuring temperature is varied, it is found that the critical field $H_C$ changes as shown in the main panel of Fig. 3(c). The evolution of the critical field $H_C$ with the different measuring temperatures is shown in of Fig. 3 (d), indicating the monotonous reduction of $H_C$ with increasing measuring temperatures. The magnetic switching effect disappears when the measuring temperature is close to the $T_N$. We try to further understand the possible magnetic ground state of CrPS$_4$ single crystal and compare the *M(H)* at *T*=5 K and *H*>*H$_C$* along different directions. As shown in Fig. 3 (a) and (b), above $H_C$, the values of *M(H)* along different directions are nearly the same, which may be due to the similar magnetic structure for the different directions. The similar character is also reported in EuCu$_2$As$_2$ and EuCu$_2$Sb$_2$,[16] the possible magnetic structure of CrPS$_4$ is presumed in Fig. 4 (a) and (b). Below $H_C$, as shown in Fig. 4 (a), the Cr$^{3+}$ moments are antiferromagnetically aligned in the *ab* plane and ferromagnetically aligned along the *c* axis, that is the collinear C-type AFM structure. The *M(T)* in Fig. 2 also consists with a collinear C-type AFM structure, where the easy magnetization axis is along the direction of *c* axis. When *H*> *H$_C$*, the magnetic structure is supposed to change into G-type AFM structure, as shown in Fig. 4(b). The Cr$^{3+}$ moments are antiferromagnetically aligned both in the *ab* plane and along the *c* axis. The G-type AFM structure is the stable magnetic state under the applied magnetic field. Fig. 4 (c) presents the possible magnetic phase diagram of CrPS$_4$ single crystal.

Based the Molecular Field Theory (MFT), we can estimate the exchange interactions. [18] In MFT, the value *f$_J$* arising from the exchange interactions as:

$$f_J = \frac{\theta_{pJ}}{T_{NJ}} = 0.54 \quad (H // c) \tag{4}$$

$\theta_{pJ}$ and *f$_J$* are related to the exchange interactions by the following functions:

$$T_{NJ} = -\frac{S(S+1)}{3k_B}\sum_j J_{ij} \cos\phi_{ji} \tag{5a}$$

$$\theta_{pJ} = -\frac{S(S+1)}{3k_B}\sum_j J_{ij} \tag{5b}$$

$$f_J = \frac{\sum_j J_{ij}}{\sum_j J_{ij}\cos\phi_{ji}} \tag{5c}$$

*i* means the center spin and *j* means all neighbors, while $J_{ij}$ is the respective exchange constant and $\Phi_{ij}$ is the angle between ordered moments $\mu_j$ and $\mu_i$ in the magnetic order state. The sums are around all neighbors *j* with which a central spin *i* interacts.

Based on the C-type AFM, we can calculate the exchange interactions of the CrPS$_4$ single crystal, which has fourfold in-plane nearest neighbor interaction $J_1$ ($J_a$ and $J_b$ are the magnetic interactions along the *a* and *b* axis in *ab* plane, based on the C-type AFM structure, $J_a$ and $J_b$ are AFM interaction. Because we cannot distinguish the interactions along *a* and *b* axes, $J_1$ is the average value of $J_a$ and $J_b$.) and twofold interlayer interaction $J_c$, we can obtain:

$$T_{NJ} = -\frac{S(S+1)}{3k_B}(2J_c - 4J_1) \tag{6a}$$

$$\theta_{pJ} = -\frac{S(S+1)}{3k_B}(2J_c + 4J_1) \tag{6b}$$

$$f_J = \frac{\theta_{pJ}}{T_{NJ}} = \frac{2J_1 + J_c}{2J_1 - J_c} \tag{6c}$$

We obtain $J_1$ = 0.143 meV, $J_c$ = -0.955 meV. $J_1$ is AFM interaction with positive value and $J_c$ is FM one with negative one, which are consistent with the C-type AFM structure. However, the further experiments, such as neutron scattering experiment, are really needed to determine the magnetic structure of CrPS$_4$ single crystal.

As to the transport properties of CrPS$_4$ single crystal, Figure 5 shows the temperature dependence of the resistivity in the *ab* plane for CrPS$_4$ single crystal. The *dρ/dT* is negative and CrPS$_4$ single crystal shows a semiconducting behavior. In the inset of Fig.5, *ln(ρ)* decreases almost linearly with increasing *T* from 175 K to 300 K, we can calculate the activation energy by using Arrhenius equation:[18]

$$\ln\rho = \ln\rho_0 - \frac{E_a}{k_B T} \tag{7}$$

As shown in the inset of Fig.5, the activation energy $E_a$ is about 0.166 eV.

To study the thermal property, we perform the measurement of specific heat of CrPS$_4$ single crystal. Fig.6 (a) shows the temperature dependence of the heat capacity $C_P$ for CrPS$_4$ single crystal. The $C_P(T)$ of CrPS$_4$ exhibits a anomaly around $T_N$ = 36 K. Because of the thin flake crystal with small mass, a sharp λ-type peak has not been observed around $T_N$=36 K. Since CrPS$_4$ shows insulating behavior, we can ignore the electronic contribution to the heat capacity, the $C_{mag}(T)$ can be calculated by the following equations:[18]

$$C_{V\,Debye}(T) = 9R\left(\frac{T}{\theta_D}\right)^3 \int_0^{\theta_D/T} \frac{x^4 e^x}{(e^x-1)^2}dx \tag{8a}$$

$$C_{mag}(T) = C_P(T) - nC_{V\,Debye}(T) \tag{8b}$$

while $n$ = 6 is the number of atoms per formula unit, $R$ is the molar gas constant and $\theta_D$ is the Debye temperature. As shown in Fig. 6 (b), the solid line is the calculated lattice contribution of the heat capacity. The fitted Debye temperature $\theta_D$ is about 502 K. We can get the $C_{mag}(T)$ based on the MFT [18], the heat-capacity jump at the magnetic transition can be calculated for the two possible magnetic structures: (1) the equal moment (EM) structure where the magnetic moments are the same at all sites and (2) the amplitude modulated (AM) structure where the magnetic-moment amplitude varies periodically from one site to another one. For EM structure, the jump in the heat capacity at the ordering temperature is given by [19]

$$\Delta C_{EM} = 5\frac{J(J+1)}{(2J^2+2J+1)}R \tag{9a}$$

and for AM structure,

$$\Delta C_{AM} = \frac{10}{3}\frac{J(J+1)}{(2J^2+2J+1)}R \tag{9b}$$

where $J$ is the total angular momentum and $R$ is the gas constant. By using $J=3/2$ for $Cr^{3+}$, $\Delta C_{EM}$ and $\Delta C_{AM}$ are calculated to 18.2 J/mol K and 12.2 J/mol K, respectively. Our estimated $\Delta C_{mag} \approx$ 10.8 J/mol K is close to the value of $\Delta C_{AM}$, which indicates that $CrPS_4$ possesses an AM structure. The AM structure obtained from the heat capapcity measurement is consistent with the above discussed C-type AFM magnetic structure. The magnetic contribution to the entroy $S_{mag}$ was obtained by integrating the $C_{mag}/T$ versus $T$:[18]

$$S_{mag}(T) = \int_0^T \frac{C_{mag}(T)}{T}dT \tag{10}$$

The $T$ dependence of $S_{mag}$ is shown in Fig. 6 (c) for the temperature range from 5 K to 100 K. The theoretical expected value of the magnetic contribution to the entropy is $S_{mag}(T) = Rln(2S+1) = Rln4 =$11.5 J/mol K (for $Cr^{3+}$, $S = 3/2$), which is shown by the dashed curve in Fig.6 (c). Fig.6 (c) indicates that the value of $S_{mag}$ is about 85 % of the theoretical expected value.

In order to investigate further the magnetic property of $CrPS_4$ single crystal, we perform its ESR spectra measurement. Fig. 7 (a) and (b) shows the ESR spectra of $CrPS_4$ singel crystal at selected temperatures under the applied field from 0 to 0.6 T, which is applied along the different crystallographic axes. The signal consists of a single exchange-narrowed resonance line, which can be well fitted by a Lorentz shape at resonance field $H_r$ with the linewidth $\Delta H$ and the half width at half maximum as shown in Fig.8 (a) and (b). The function of Lorentz line is [20]:

$$Y' = -Y_{max}\frac{2\Delta H^2(H-H_r)}{[\Delta H^2+(H-H_r)^2]^2} \tag{11a}$$

$$Y = Y_{max}\frac{\Delta H^2}{\Delta H^2+(H-H_r)^2} \tag{11b}$$

$$Y_{max} = Y|_{H=H_r} \tag{11c}$$

The temperature dependence of the effective $g$ values along different directions are plotted in Fig. 8 (c) and (d). The $g$ values at different temperatures are obtained from resonance field $H_r$ and the microwave frequency $\nu$ via the Larmor condition [20]:

$$h\nu = g\mu_B H_r \tag{12}$$

The calculated $g$ values are around 1.98, which is in good agreement with the previously reported result of $g_{Cr^{3+}} = 1.978$ [21].

Figure 8 (e) and (f) display the normalized ESR intensity $I$ by its value at 300 K as a function of temperature. We obtain the value of $I$ according to the following function:

$$I = \int_{+\infty}^{0} Y\, dH \tag{13}$$

which can be described by a thermally activated model, and well fitted by the Arrhenius law [22]:

$$I \propto exp(\Delta E/k_B T) \tag{14}$$

where $\Delta E$ is the activation energy. The insets of Fig.8 (e) and (f) show $ln\ I/I$(T=300 K) vs. 1000/T. According to Eq. (14), we can obtain the activation energy $\Delta E(H /\!/ c)$=0.0135 eV and $\Delta E(H \perp c)$=0.0144 eV.

The angular dependence of the ESR linewidth includes the more information on the exchange interaction. As shown in Fig. (9), we investigate the angular dependence of the ESR linewidth in detail for three crystallographic planes at $T$=150 K. The three lines are coresponding to a set of orthogonal axes: the black line is coresponding to the applied field $H$ perpendicular to the $c$ axis and red lines are coresponding to a pair of orthogonal axes lying in the $ab$ plane.

The ESR linewidth $\Delta H$ can be fitted via the relation between the second moment $M_2$ [23, 24]:

$$\Delta H = \frac{\hbar}{g\mu_B \omega_{ex}} M_2 \tag{15}$$

where $\omega_{ex}$ is so-called exchange frequency. In this article, we use the experimental value of $g = 1.98$, which is obtained from the ESR measurement.

The second moment $M_2$ is given by:

$$M_2 = 2\frac{S(S+1)}{3}\left\{f_1\left(2\tilde{J}_{zz} - \tilde{J}_{xx} - \tilde{J}_{yy}\right)^2 + f_2 \times 10\left(\tilde{J}_{xz}^2 + \tilde{J}_{yz}^2\right) + f_3\left[\left(\tilde{J}_{xx} - \tilde{J}_{yy}\right)^2 + 4\tilde{J}_{xy}^2\right]\right\} \tag{16}$$

The symbols $f_1$, $f_2$ and $f_3$ denote the spectral-density functions. [23, 24]

The angular dependence of the ESR linewidth at $T$=150 K can be well fitted. The fitting solid lines are illustrated in Fig. (9). The obtained parameters are following: $J_1/J_c$ = -0.635 with $J_1 > 0$ and $J_c < 0$. $J_1 > 0$ means AFM interaction in $ab$ plane and $J_c < 0$ means FM interaction interlayer, which are again in agreement with the C-type AFM structure discussed above. However, the value of $J_1/J_c$ is smaller than the value obtained from the magnetization result mentioned above. The framework of our model is simple and does not include other interactions. For example, Dzyaloshinsky-Moriya (DM) interaction may result in the tilting of the $CrS_6$ octahedra along the antiferromagnetically coupled along *c* axis, which can induce line broadening of comparable order of magnitude.[25]

As it is well known, many chalcogenides form layered structures and can be exfoliated, the complexity of bulk crystals can be well tuned in the single- and few-layer materials. Recently, monolayer of the Cr-based tri-chalcogenides, such as Cr(Si, Ge and Sn)$Te_3$, have been well studied by the first principle calculations, which show that the physical properties of monolayer Cr-based tri- chalcogenides is very sensitive to the strain effecting.[15, 26, 27] For example, for $CrSiTe_3$, with the help of the strain, the ground state can be well tuned from AFM semiconductor to FM one. The FM Curie temperature of the monolayer $CrSiTe_3$ can be up to 156 K. Herein, we compare the exchange interactions between the 2D $AMX_3$ compounds and $CrPS_4$, as shown in **Table IV**.[28] The bulk $CrPS_4$ single crystal shows a similar character with the 2D $AMX_3$ materials. That means with the application of a moderate uniform in-plane tensile strain of 3-4 %, which is experimentally feasible, the ground state may be tunable from AFM semiconductor of bulk $CrPS_4$ into FM one in monolayer $CrPS_4$. On the other hand, compared with Cr(Si, Ge and Sn)$Te_3$, there are some advantages for $CrPS_4$ single crystal. One is that the monolayer of $CrPS_4$ can be easily obtained by exfoliating because $CrPS_4$ single crystal is soft. Another advantage is that the $CrPS_4$ single crystal is very stable in air, which is good for fabricating the heterojunction with other chalcogenides in the spintronic application. Therefore, the monolayer $CrPS_4$ may be a promising candidate to explore 2D magnetic semiconductors. However, it is still an open question, further research, such as the first-principle calculations or the investigation on the mono-layer $CrPS_4$ crystal or $CrPS_4$ film are really needed.

## IV Conclusion

The detailed physical properties of the layered compound $CrPS_4$ single crystal were investigated by the room-temperature X-ray diffraction, electrical resistivity $\rho(T)$, and specific heat $C_P(T)$. $CrPS_4$ crystals crystallize into a monoclinic structure. The electrical resistivity $\rho(T)$ shows an insulator behavior with an energy gap $E_a$=0.167 eV. The AFM transition temperature is about $T_N$=36 K. The exchange constants $J_1$ and $J_c$ corresponding to the intralayer and interlayer interaction are about 0.143 *meV* and -0.955 *meV*. The magneto-switching effect is obtained along the *c* axis, which may be related to the change of magnetic structure induced by the applied magnetic field. The possible magnetic phase diagram of CrPS$_4$ single crystal has been presented. The extracted magnetic entropy at $T_N$ is consistent with the theoretical value $R \ln(2S + 1)$ for $S = 3/2$ of the Cr$^{3+}$ ion. It is suggested that monolayer CrPS$_4$ may be a promising candidate to explore 2D magnetic semiconductors with the help of the strain effect.

## Acknowledgement

This work was supported by the Joint Funds of the National Natural Science Foundation of China and the Chinese Academy of Sciences' Large-Scale Scientific Facility under contracts (U1432139, U1232139), the National Nature Science Foundation of China under contracts (51171177, 11404342), the National Key Basic Research under contract 2011CBA00111, and the Nature Science Foundation of Anhui Province under contract 1508085ME103, 1408085MA11.

**Table I**: Crystallographic parameters of CrPS$_4$ single crystal.

| Atom | Wyckoff position | $x$ | $y$ | $z$ |
|------|------------------|--------|--------|--------|
| Cr | 4g | 0 | 0.2481 | 0 |
| P | 4i | 0.2971 | 0 | 0.1655 |
| S1 | 8j | 0.1341 | 0.2666 | 0.7 |
| S2 | 4i | 0.1062 | 0 | 0.1934 |
| S3 | 4i | 0.1303 | 0.5 | 0.1452 |

**Table II:** Crystallographic and Rietveld refinement obtained from powder XRD data.

| Crystal system | hexagonal |
|---|---|
| Space group | $C\ 1\ 2/m\ 1$ (No.12) |
| Wavelength | 1.5406 (*Cu K$_\alpha$*) |
| Unit cell | |
| $a$ (Å) | 10.871 |
| $b$ (Å) | 7.254 |
| $c$ (Å) | 6.14 |
| $\beta$ (º) | 91.88 |
| V (Å$^3$) | 483.929 |
| Z | 4 |
| $d_{Cr-S}$ (Å) | 2.39 |
| | 2.4275 |
| | 2.4611 |
| $d_{P-S}$ (Å) | 2.0168 |
| | 2.0879 |
| | 2.0883 |
| $\rho_{calc}$ (g/cm$^3$) | 2.9 |
| $\chi^2$ | 2.866 |
| $R_p$ | 9.414 |
| $R_{wp}$ | 8.211 |

**Table III**: The physical parameters for CrPS$_4$ single crystal.

| parameters | CrPS$_4$ |
|---|---|
| $T_N$ (K) | 36 |
| $\mu_{eff}$ | 6.47 $\mu_0$ ($H /\!/ c$) |
| | 6.41 $\mu_0$ ($H \perp c$) |
| $\Theta$ (K) | 19.4 |
| $J_1$ (meV) | 0.143 |
| $J_C$ (meV) | -0.955 |
| $\rho(300\ K)$ ($\Omega$ cm) | 1.2×10$^3$ |
| $E_a$ (eV) | 0.166 |
| $\Theta_D$ (K) | 502 |
| $\Delta C_{mag}$ (J/mol K) | 10.8 |
| $S_{mag}$ (J/mol K) | 11.5 |
| $\Delta E$ (eV) | 0.0135 ($H /\!/ c$) |
| | 0.0144 ($H \perp c$) |

**Table IV**: Lattice constant $a$, magnetic ground state and magnetic critical temperature for reported ABX$_3$ and CrPS$_4$.

| | $a$ (Å) | Ground state | $J$ (meV) | $T_C$ ($T_N$) (K) | Ref. |
|---|---|---|---|---|---|
| MnPSe$_3$ | 6.27 | AFM-Néel | 0.46 | 147 | 28 |
| MnPSe$_3$ (2 % strain) | 6.40 | AFM-Néel | 0.33 | 115 | 28 |
| CrSiTe$_3$ | 6.84 | AFM-zigzag | -0.74 | 160 | 28 |
| CrSiTe$_3$ (4% strain) | 7.11 | FM | -2.29 | 158 | 28 |
| CrPS$_4$ | 10.871 | C-type AFM | 0.143 | 36 | this work |

**Figure 1:**

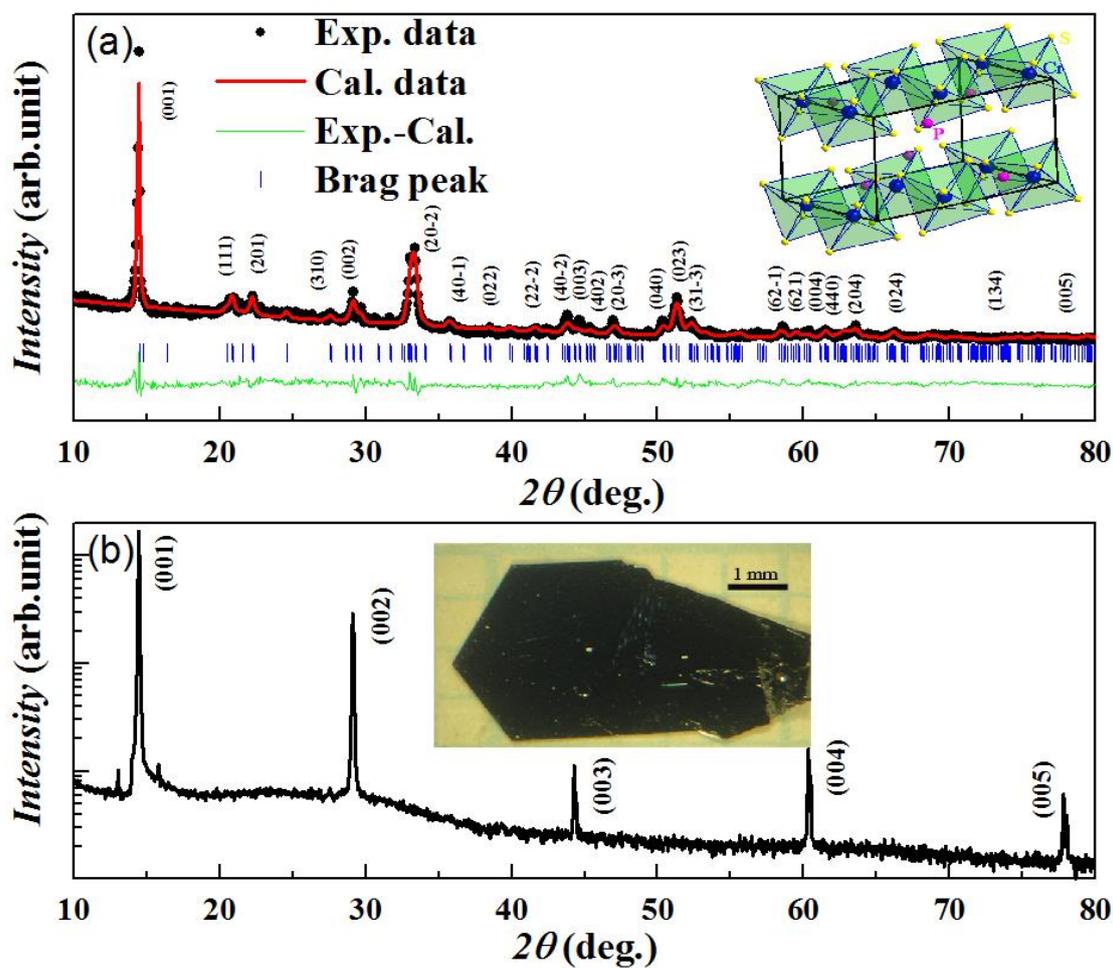

**Fig. 1(color online):** (a) The Rietveld refined powder XRD patterns at room temperature for the crushed $CrPS_4$ crystals. The vertical marks (blue bars) stand for the position of Brag peaks, and solid line (green line) at the bottom correspond to the difference between experimental and calculated intensities. Inset shows the crystal structure of $CrPS_4$ drawn based on the XRD refinement. (b) XRD patterns of the crystal measured on the (001) surface. Inset presents the picture of the $CrPS_4$ single crystal used for this study. The crystal size is approximately 4*6*0.5 mm.

**Figure 2:**

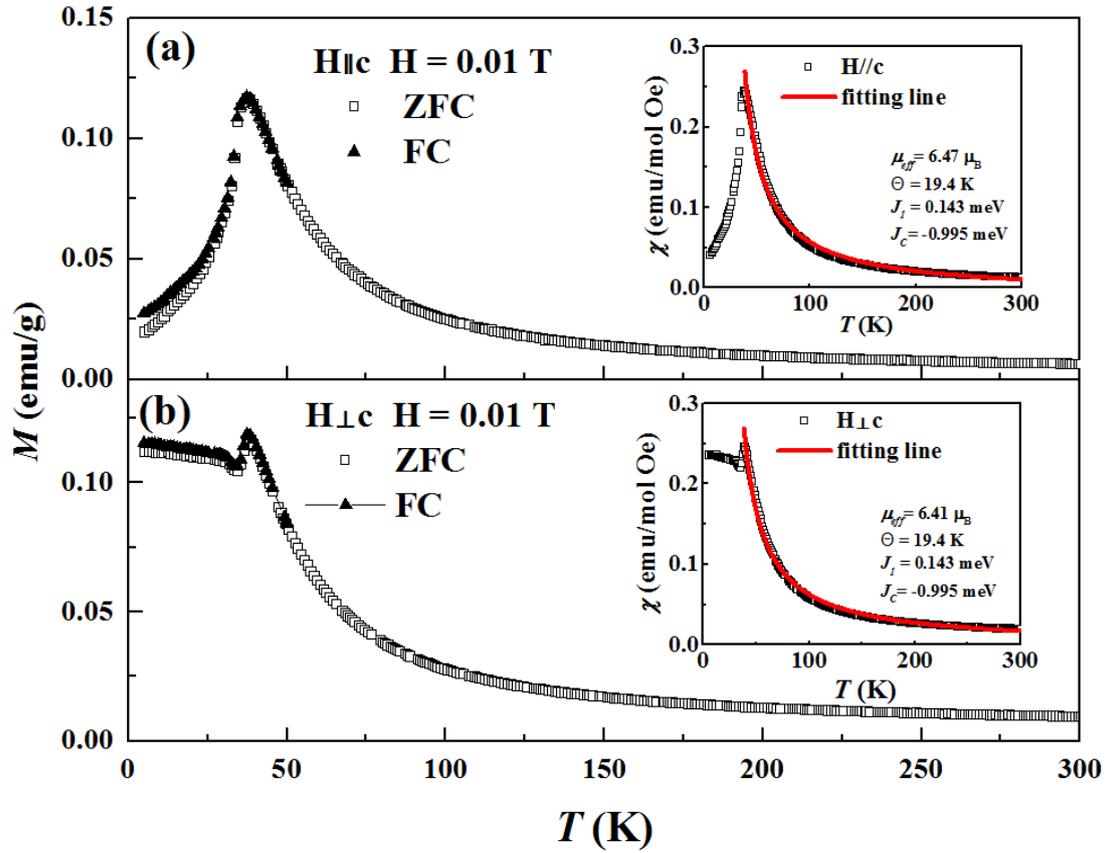

**Fig. 2 (color online):** The temperature dependence of magnetization of CrPS$_4$ single crystal under ZFC and FC models with applied magnetic field H=0.01 T. (a) for applied magnetic field parallel to the *c* axis and (b) for the applied magnetic field perpendicular to the *c* axis. The insets of (a) and (b) show the fitting results according to the Eq. (1).

**Figure 3:**

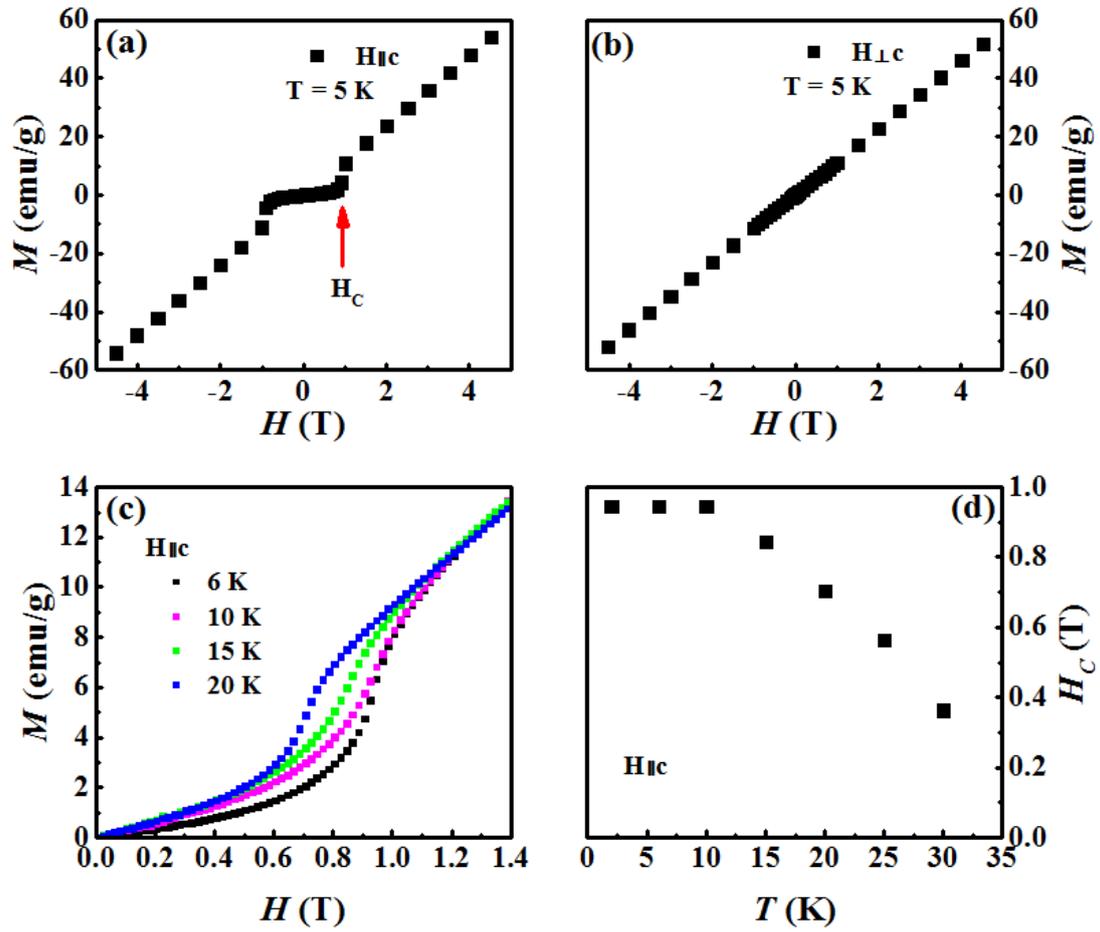

**Fig. 3 (color online):** (a) and (b) show the magnetic field dependence of magnetization at $T$=5 K for $H$ parallel and perpendicular to the $c$ axis, respectively. (c) shows the magnetic field dependence of magnetization at different temperatures. (d) presents the temperature dependence of critical field ($H_C$).

**Figure 4:**

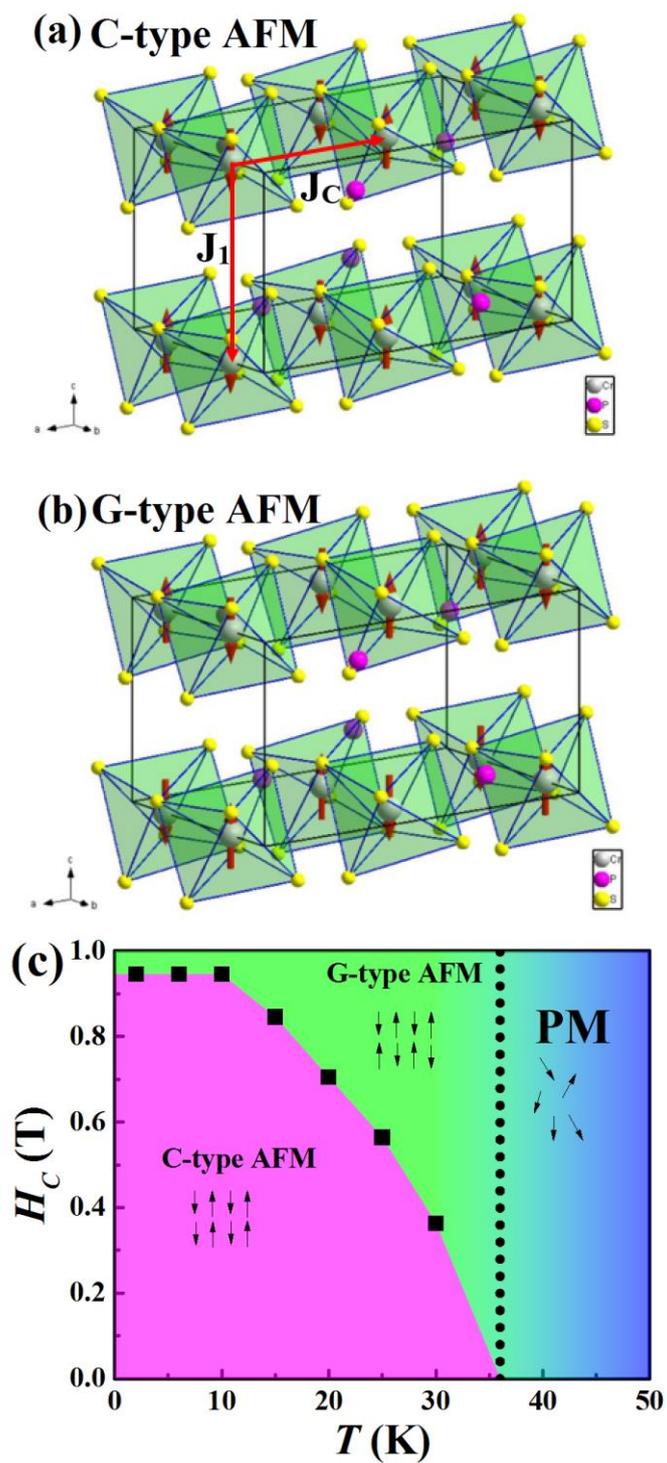

**Fig. 4 (color online):** The magnetic structures of CrPS$_4$ (a) the C-type AFM below the critical magnetization $H_C$; (b) G-type AFM for the magnetic structure above the $H_C$ and (c) The magnetic phase diagram of CrPS$_4$.

**Figure 5:**

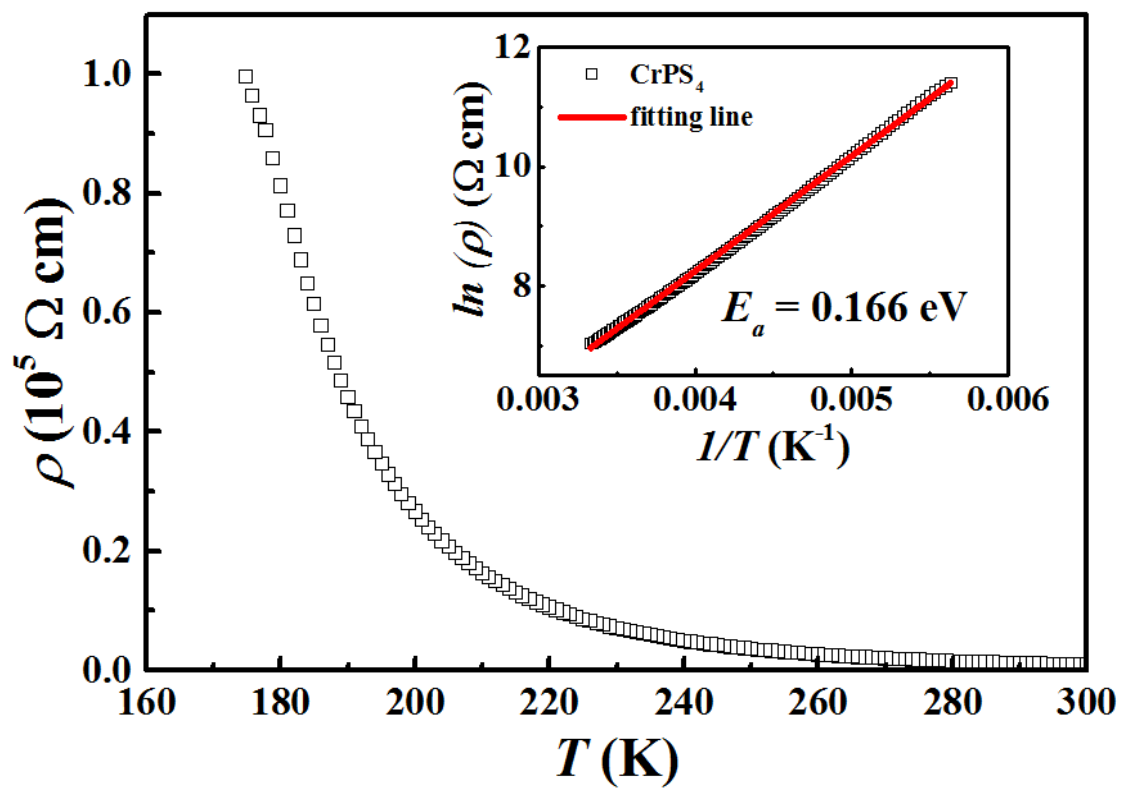

**Fig. 5 (color online):** The temperature dependence of resistivity. The inset shows the fitting results according to the thermal active model.

**Figure 6:**

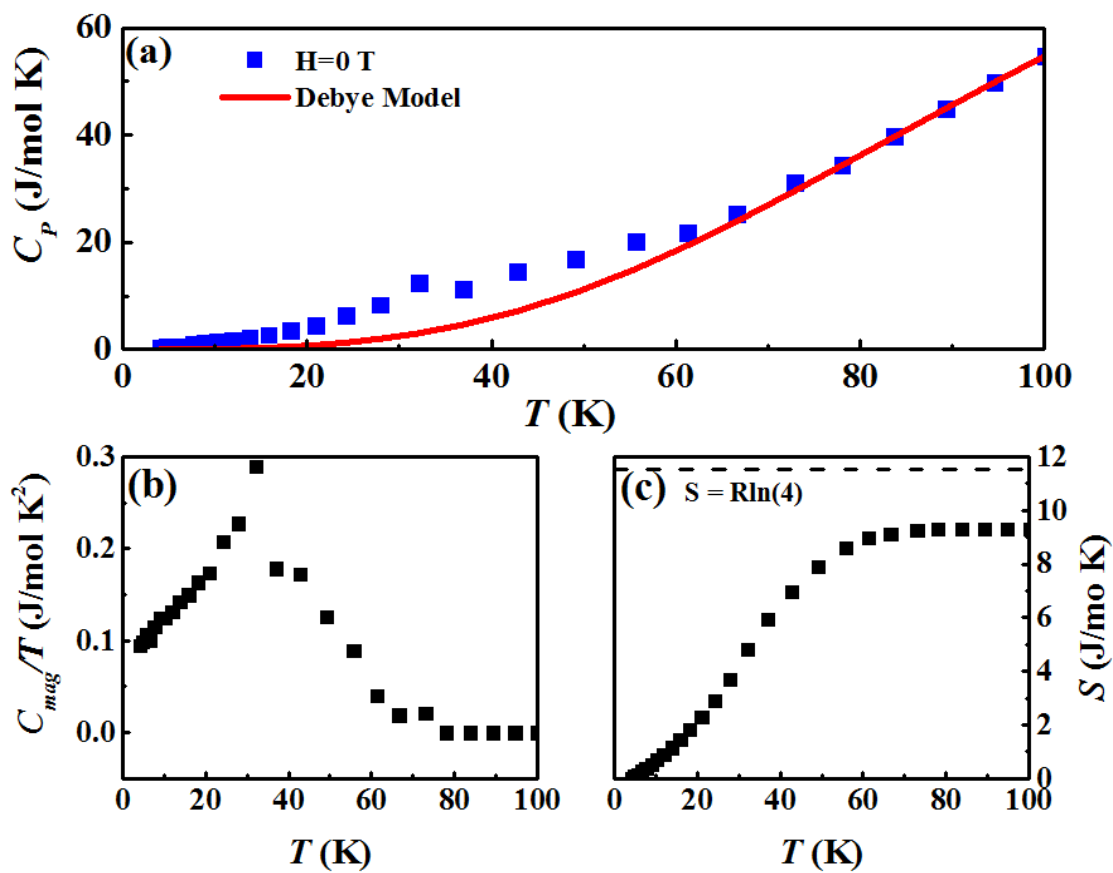

**Fig. 6 (color online):** (a) Heat capacity as a function of temperature under the applied magnetic field $H=0$ T. The solid line is the fitting line according to the Debye model. (b) $C_{mag}/T$ vs. $T$. (c) The calculated magnetic entropy $S_{mag}$ vs. $T$ of CrPS$_4$ single crystal.

**Figure 7:**

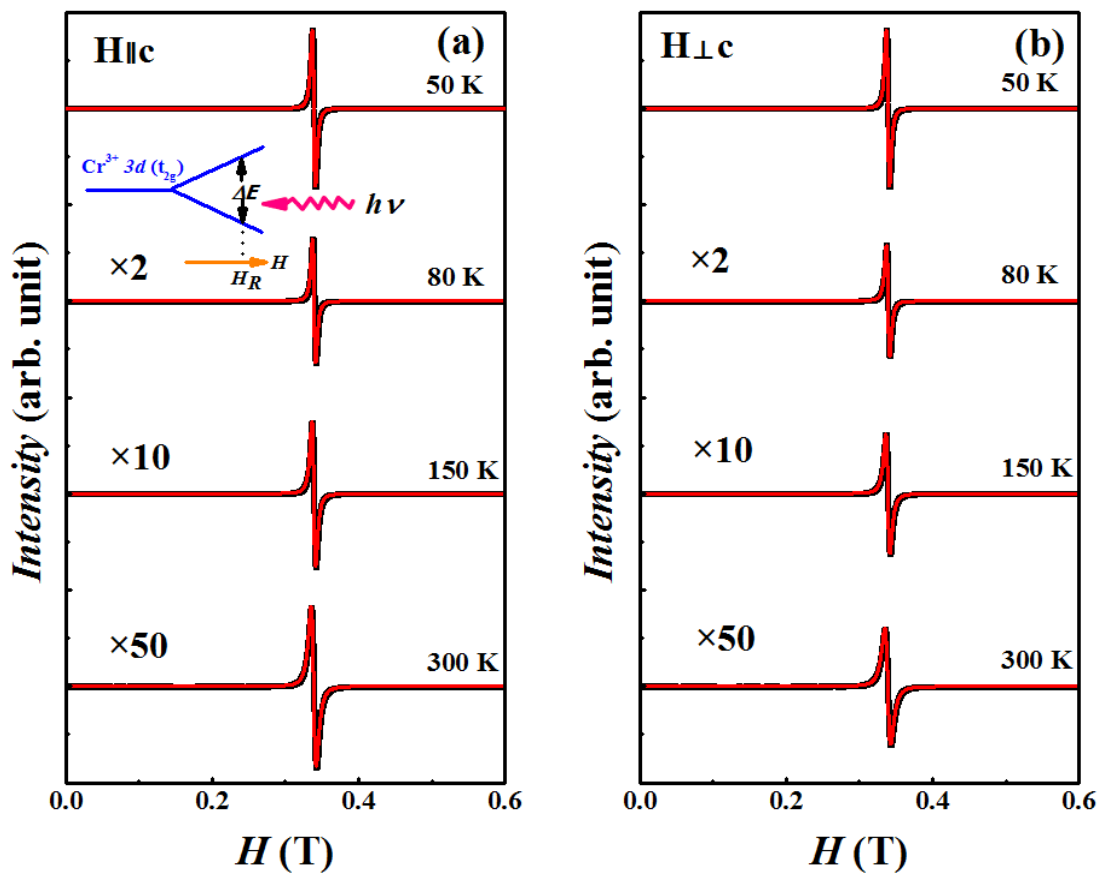

**Fig. 7 (color online):** The magnetic field dependence of ESR spectra along the different crystallographic axes. (a) for magnetic field parallel to the *c* axis and (b) for magnetic field perpendicular to the *c* axis. The red line is the fitting line according to a Lorentz shape.

**Figure 8:**

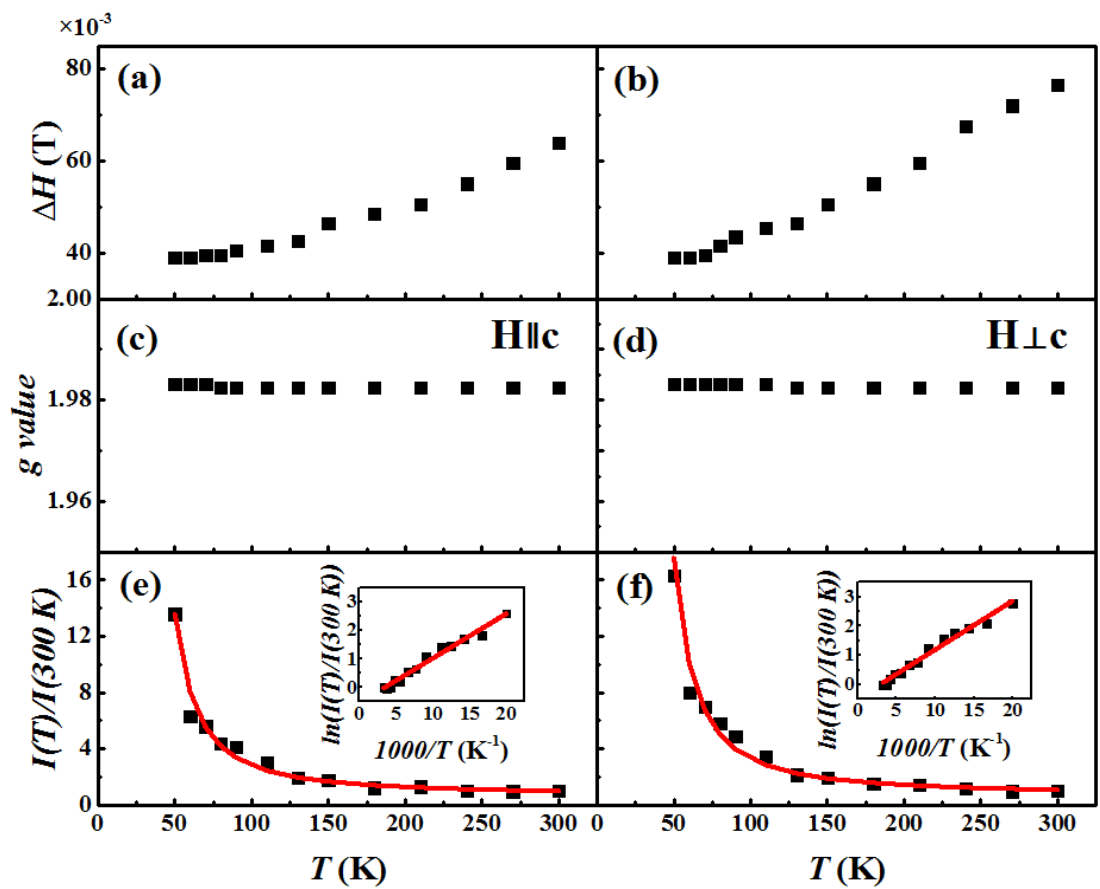

**Fig. 8 (color online):** (a) (b) The temperature dependence of HWHM (the half width at half maximum) $\Delta H$ along different crystallographic axes. (c) (d) The temperature dependence of the effective *g* values along different crystallographic axes. (e) (f) The normalized ESR intensity *I* by its value at 300 K *vs. T* of CrPS$_4$ single crystal. The red lines and the inset pictures show the fitting results according to the thermally activated model.

**Figure 9:**

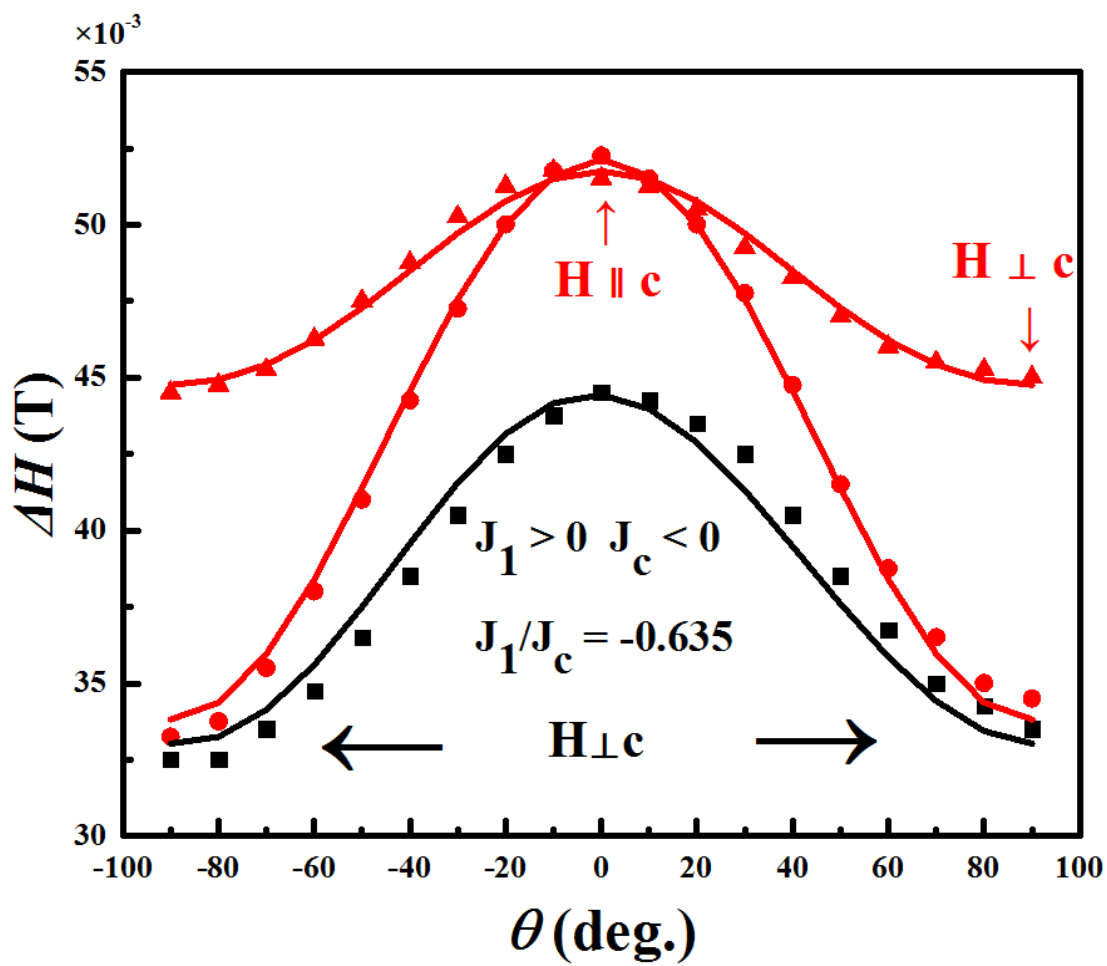

**Fig. 9 (color online):** Angular dependence of the HWHM (the half width at half maximum) $\varDelta H$ and $H_{res}$ for three crystallographic planes in different degrees. The solid lines show the fitting results of in-plane interaction and interlayer one.